
%
%
%
%
%
%
%
%
\def\standardrisposta{s }\def\reducedrisposta{r }
\def\mplarisposta{mpla }\def\zerorisposta{z }
\def\doublerisposta{d }\def\cartarisposta{e }\def\amsrisposta{y }
\newcount\ingrandimento \newcount\sinnota \newcount\dimnota
\newcount\unoduecol \newdimen\collhsize \newdimen\tothsize
\newdimen\fullhsize \newcount\controllorisposta \sinnota=1
\newskip\infralinea  \global\controllorisposta=0
\immediate\write16 { ********  Welcome to PANDA macros (Plain TeX,
AP, 1991) ******** }
\immediate\write16 { You'll have to answer a few questions in
lowercase.}
\message{>  Do you want it in double-page (d), reduced (r)
or standard format (s) ? }\read-1 to\risposta
\message{>  Do you want it in USA A4 (u) or EUROPEAN A4
(e) paper size ? }\read-1 to\srisposta
\message{>  Do you have AMSFonts 2.0 (math) fonts (y/n) ? }
\read-1 to\arisposta
%
%
%
%
%
\ifx\risposta\standardrisposta \ingrandimento=1200
\message {>> This will come out UNREDUCED << }
\dimnota=2 \unoduecol=1 \global\controllorisposta=1 \fi
\ifx\risposta\reducedrisposta \ingrandimento=1095 \dimnota=1
\unoduecol=1  \global\controllorisposta=1
\message {>> This will come out REDUCED << } \fi
\ifx\risposta\doublerisposta \ingrandimento=1000 \dimnota=2
\unoduecol=2  \message {>> You must print this in
LANDSCAPE orientation << } \global\controllorisposta=1 \fi
\ifx\risposta\mplarisposta \ingrandimento=1000 \dimnota=1
\message {>> Mod. Phys. Lett. A format << }
\unoduecol=1 \global\controllorisposta=1 \fi
\ifx\risposta\zerorisposta \ingrandimento=1000 \dimnota=2
\message {>> Zero Magnification format << }
\unoduecol=1 \global\controllorisposta=1 \fi
\ifnum\controllorisposta=0  \ingrandimento=1200
\message {>>> ERROR IN INPUT, I ASSUME STANDARD
UNREDUCED FORMAT <<< }  \dimnota=2 \unoduecol=1 \fi
\magnification=\ingrandimento
%
%
%
%
\newdimen\eucolumnsize \newdimen\eudoublehsize \newdimen\eudoublevsize
\newdimen\uscolumnsize \newdimen\usdoublehsize \newdimen\usdoublevsize
\newdimen\eusinglehsize \newdimen\eusinglevsize \newdimen\ussinglehsize
\newskip\standardbaselineskip \newdimen\ussinglevsize
\newskip\reducedbaselineskip \newskip\doublebaselineskip
\eucolumnsize=12.0truecm    
\eudoublehsize=25.5truecm   
\eudoublevsize=6.5truein    
\uscolumnsize=4.4truein     
\usdoublehsize=9.4truein    
\usdoublevsize=6.8truein    
\eusinglehsize=6.5truein    
\eusinglevsize=24truecm     
\ussinglehsize=6.5truein    
\ussinglevsize=8.9truein    
\standardbaselineskip=16pt plus.2pt  
\reducedbaselineskip=14pt plus.2pt   
\doublebaselineskip=12pt plus.2pt    
%
%
\def\Portoffset{}
\def\Landoffset{}
\ifx\risposta\mplarisposta \def\Portoffset{\hoffset=1.8truecm} \fi
%
%
\def\Landspec{}
\tolerance=10000
\parskip=0pt plus2pt  \leftskip=0pt \rightskip=0pt
%
%
\ifx\risposta\standardrisposta \infralinea=\standardbaselineskip \fi
\ifx\risposta\reducedrisposta  \infralinea=\reducedbaselineskip \fi
\ifx\risposta\doublerisposta   \infralinea=\doublebaselineskip \fi
\ifx\risposta\mplarisposta     \infralinea=13pt \fi
\ifx\risposta\zerorisposta     \infralinea=12pt plus.2pt\fi
\ifnum\controllorisposta=0    \infralinea=\standardbaselineskip \fi
\ifx\risposta\doublerisposta   \Landoffset \else \Portoffset \fi
\ifx\risposta\doublerisposta \ifx\srisposta\cartarisposta
\tothsize=\eudoublehsize \collhsize=\eucolumnsize
\vsize=\eudoublevsize  \else  \tothsize=\usdoublehsize
\collhsize=\uscolumnsize \vsize=\usdoublevsize \fi \else
\ifx\srisposta\cartarisposta \tothsize=\eusinglehsize
\vsize=\eusinglevsize \else  \tothsize=\ussinglehsize
\vsize=\ussinglevsize \fi \collhsize=4.4truein \fi
\ifx\risposta\mplarisposta \tothsize=5.0truein
\vsize=7.8truein \collhsize=4.4truein \fi
%
%
%
%
\newcount\contaeuler \newcount\contacyrill \newcount\contaams
\font\ninerm=cmr9  \font\eightrm=cmr8  \font\sixrm=cmr6
\font\ninei=cmmi9  \font\eighti=cmmi8  \font\sixi=cmmi6
\font\ninesy=cmsy9  \font\eightsy=cmsy8  \font\sixsy=cmsy6
\font\ninebf=cmbx9  \font\eightbf=cmbx8  \font\sixbf=cmbx6
\font\ninett=cmtt9  \font\eighttt=cmtt8  \font\nineit=cmti9
\font\eightit=cmti8 \font\ninesl=cmsl9  \font\eightsl=cmsl8
\skewchar\ninei='177 \skewchar\eighti='177 \skewchar\sixi='177
\skewchar\ninesy='60 \skewchar\eightsy='60 \skewchar\sixsy='60
\hyphenchar\ninett=-1 \hyphenchar\eighttt=-1 \hyphenchar\tentt=-1
%
\font\tencmmib=cmmib10  \newfam\cmmibfam  \skewchar\tencmmib='177
\font\tencmbsy=cmbsy10  \newfam\cmbsyfam  \skewchar\tencmbsy='60
\font\tencmcsc=cmcsc10  \newfam\cmcscfam
\ifnum\ingrandimento=1095

\font\capsone=cmcsc10 at 10.95pt 

\else

\font\capsone=cmcsc10 at 12pt 
\fi

\def\ttaarr{\bf}		
\def\ppaarr{\sl}		

%
%
%
\newfam\eufmfam \newfam\msamfam \newfam\msbmfam \newfam\eufbfam
\def\Loadeulerfonts{\global\contaeuler=1 \ifx\arisposta\amsrisposta
\font\teneufm=eufm10              
\font\eighteufm=eufm8 \font\nineeufm=eufm9 \font\sixeufm=eufm6
\font\seveneufm=eufm7  \font\fiveeufm=eufm5
\font\teneufb=eufb10              
\font\eighteufb=eufb8 \font\nineeufb=eufb9 \font\sixeufb=eufb6
\font\seveneufb=eufb7  \font\fiveeufb=eufb5
\font\teneurm=eurm10              
\font\eighteurm=eurm8 \font\nineeurm=eurm9
\font\teneurb=eurb10              
\font\eighteurb=eurb8 \font\nineeurb=eurb9
\font\teneusm=eusm10              
\font\eighteusm=eusm8 \font\nineeusm=eusm9
\font\teneusb=eusb10              
\font\eighteusb=eusb8 \font\nineeusb=eusb9
\else \def\eufm{\tt} \def\eufb{\tt} \def\eurm{\tt} \def\eurb{\tt}
\def\eusm{\tt} \def\eusb{\tt}    \fi}

\def\loadamsmath{\global\contaams=1 \ifx\arisposta\amsrisposta
\font\tenmsam=msam10 \font\ninemsam=msam9 \font\eightmsam=msam8
\font\sevenmsam=msam7 \font\sixmsam=msam6 \font\fivemsam=msam5
\font\tenmsbm=msbm10 \font\ninemsbm=msbm9 \font\eightmsbm=msbm8
\font\sevenmsbm=msbm7 \font\sixmsbm=msbm6 \font\fivemsbm=msbm5
\else \def\msbm{\bf} \fi \def\Bbb{\msbm} \def\symbl{\msam} \tenpoint}
\def\loadcyrill{\global\contacyrill=1 \ifx\arisposta\amsrisposta
\font\tenwncyr=wncyr10 \font\ninewncyr=wncyr9 \font\eightwncyr=wncyr8
\font\tenwncyb=wncyr10 \font\ninewncyb=wncyr9 \font\eightwncyb=wncyr8
\font\tenwncyi=wncyr10 \font\ninewncyi=wncyr9 \font\eightwncyi=wncyr8
\else \def\cyrill{\sl} \def\cyrilb{\sl} \def\cyrili{\sl} \fi\tenpoint}
\ifx\arisposta\amsrisposta
\font\sevenex=cmex7               
\font\eightex=cmex8  \font\nineex=cmex9
\font\ninecmmib=cmmib9   \font\eightcmmib=cmmib8
\font\sevencmmib=cmmib7 \font\sixcmmib=cmmib6
\font\fivecmmib=cmmib5   \skewchar\ninecmmib='177
\skewchar\eightcmmib='177  \skewchar\sevencmmib='177
\skewchar\sixcmmib='177   \skewchar\fivecmmib='177
\font\ninecmbsy=cmbsy9    \font\eightcmbsy=cmbsy8
\font\sevencmbsy=cmbsy7  \font\sixcmbsy=cmbsy6
\font\fivecmbsy=cmbsy5   \skewchar\ninecmbsy='60
\skewchar\eightcmbsy='60  \skewchar\sevencmbsy='60
\skewchar\sixcmbsy='60    \skewchar\fivecmbsy='60
\font\ninecmcsc=cmcsc9    \font\eightcmcsc=cmcsc8     \else
\def\cmmib{\fam\cmmibfam\tencmmib}\textfont\cmmibfam=\tencmmib
\scriptfont\cmmibfam=\tencmmib \scriptscriptfont\cmmibfam=\tencmmib
\def\cmbsy{\fam\cmbsyfam\tencmbsy} \textfont\cmbsyfam=\tencmbsy
\scriptfont\cmbsyfam=\tencmbsy \scriptscriptfont\cmbsyfam=\tencmbsy
\scriptfont\cmcscfam=\tencmcsc \scriptscriptfont\cmcscfam=\tencmcsc
\def\cmcsc{\fam\cmcscfam\tencmcsc} \textfont\cmcscfam=\tencmcsc \fi
\catcode`@=11
\newskip\ttglue
\gdef\tenpoint{\def\rm{\fam0\tenrm}
  \textfont0=\tenrm \scriptfont0=\sevenrm \scriptscriptfont0=\fiverm
  \textfont1=\teni \scriptfont1=\seveni \scriptscriptfont1=\fivei
  \textfont2=\tensy \scriptfont2=\sevensy \scriptscriptfont2=\fivesy
  \textfont3=\tenex \scriptfont3=\tenex \scriptscriptfont3=\tenex
  \def\mcal{\fam2 \tensy}  \def\mmit{\fam1 \teni}
  \textfont\itfam=\tenit \def\it{\fam\itfam\tenit}
  \textfont\slfam=\tensl \def\sl{\fam\slfam\tensl}
  \textfont\ttfam=\tentt \scriptfont\ttfam=\eighttt
  \scriptscriptfont\ttfam=\eighttt  \def\tt{\fam\ttfam\tentt}
  \textfont\bffam=\tenbf \scriptfont\bffam=\sevenbf
  \scriptscriptfont\bffam=\fivebf \def\bf{\fam\bffam\tenbf}
     \ifx\arisposta\amsrisposta    \ifnum\contaeuler=1
  \textfont\eufmfam=\teneufm \scriptfont\eufmfam=\seveneufm
  \scriptscriptfont\eufmfam=\fiveeufm \def\eufm{\fam\eufmfam\teneufm}
  \textfont\eufbfam=\teneufb \scriptfont\eufbfam=\seveneufb
  \scriptscriptfont\eufbfam=\fiveeufb \def\eufb{\fam\eufbfam\teneufb}
  \def\eurm{\teneurm} \def\eurb{\teneurb} \def\eusm{\teneusm}
  \def\eusb{\teneusb}    \fi    \ifnum\contaams=1
  \textfont\msamfam=\tenmsam \scriptfont\msamfam=\sevenmsam
  \scriptscriptfont\msamfam=\fivemsam \def\msam{\fam\msamfam\tenmsam}
  \textfont\msbmfam=\tenmsbm \scriptfont\msbmfam=\sevenmsbm
  \scriptscriptfont\msbmfam=\fivemsbm \def\msbm{\fam\msbmfam\tenmsbm}
     \fi      \ifnum\contacyrill=1     \def\cyrill{\tenwncyr}
  \def\cyrilb{\tenwncyb}  \def\cyrili{\tenwncyi}         \fi
  \textfont3=\tenex \scriptfont3=\sevenex \scriptscriptfont3=\sevenex
  \def\cmmib{\fam\cmmibfam\tencmmib} \scriptfont\cmmibfam=\sevencmmib
  \textfont\cmmibfam=\tencmmib  \scriptscriptfont\cmmibfam=\fivecmmib
  \def\cmbsy{\fam\cmbsyfam\tencmbsy} \scriptfont\cmbsyfam=\sevencmbsy
  \textfont\cmbsyfam=\tencmbsy  \scriptscriptfont\cmbsyfam=\fivecmbsy
  \def\cmcsc{\fam\cmcscfam\tencmcsc} \scriptfont\cmcscfam=\eightcmcsc
  \textfont\cmcscfam=\tencmcsc \scriptscriptfont\cmcscfam=\eightcmcsc
     \fi            \tt \ttglue=.5em plus.25em minus.15em
  \normalbaselineskip=12pt
  \setbox\strutbox=\hbox{\vrule height8.5pt depth3.5pt width0pt}
  \let\sc=\eightrm \let\big=\tenbig   \normalbaselines
  \baselineskip=\infralinea  \rm}
\gdef\ninepoint{\def\rm{\fam0\ninerm}
  \textfont0=\ninerm \scriptfont0=\sixrm \scriptscriptfont0=\fiverm
  \textfont1=\ninei \scriptfont1=\sixi \scriptscriptfont1=\fivei
  \textfont2=\ninesy \scriptfont2=\sixsy \scriptscriptfont2=\fivesy
  \textfont3=\tenex \scriptfont3=\tenex \scriptscriptfont3=\tenex
  \def\mcal{\fam2 \ninesy}  \def\mmit{\fam1 \ninei}
  \textfont\itfam=\nineit \def\it{\fam\itfam\nineit}
  \textfont\slfam=\ninesl \def\sl{\fam\slfam\ninesl}
  \textfont\ttfam=\ninett \scriptfont\ttfam=\eighttt
  \scriptscriptfont\ttfam=\eighttt \def\tt{\fam\ttfam\ninett}
  \textfont\bffam=\ninebf \scriptfont\bffam=\sixbf
  \scriptscriptfont\bffam=\fivebf \def\bf{\fam\bffam\ninebf}
     \ifx\arisposta\amsrisposta  \ifnum\contaeuler=1
  \textfont\eufmfam=\nineeufm \scriptfont\eufmfam=\sixeufm
  \scriptscriptfont\eufmfam=\fiveeufm \def\eufm{\fam\eufmfam\nineeufm}
  \textfont\eufbfam=\nineeufb \scriptfont\eufbfam=\sixeufb
  \scriptscriptfont\eufbfam=\fiveeufb \def\eufb{\fam\eufbfam\nineeufb}
  \def\eurm{\nineeurm} \def\eurb{\nineeurb} \def\eusm{\nineeusm}
  \def\eusb{\nineeusb}     \fi   \ifnum\contaams=1
  \textfont\msamfam=\ninemsam \scriptfont\msamfam=\sixmsam
  \scriptscriptfont\msamfam=\fivemsam \def\msam{\fam\msamfam\ninemsam}
  \textfont\msbmfam=\ninemsbm \scriptfont\msbmfam=\sixmsbm
  \scriptscriptfont\msbmfam=\fivemsbm \def\msbm{\fam\msbmfam\ninemsbm}
     \fi       \ifnum\contacyrill=1     \def\cyrill{\ninewncyr}
  \def\cyrilb{\ninewncyb}  \def\cyrili{\ninewncyi}         \fi
  \textfont3=\nineex \scriptfont3=\sevenex \scriptscriptfont3=\sevenex
  \def\cmmib{\fam\cmmibfam\ninecmmib}  \textfont\cmmibfam=\ninecmmib
  \scriptfont\cmmibfam=\sixcmmib \scriptscriptfont\cmmibfam=\fivecmmib
  \def\cmbsy{\fam\cmbsyfam\ninecmbsy}  \textfont\cmbsyfam=\ninecmbsy
  \scriptfont\cmbsyfam=\sixcmbsy \scriptscriptfont\cmbsyfam=\fivecmbsy
  \def\cmcsc{\fam\cmcscfam\ninecmcsc} \scriptfont\cmcscfam=\eightcmcsc
  \textfont\cmcscfam=\ninecmcsc \scriptscriptfont\cmcscfam=\eightcmcsc
     \fi            \tt \ttglue=.5em plus.25em minus.15em
  \normalbaselineskip=11pt
  \setbox\strutbox=\hbox{\vrule height8pt depth3pt width0pt}
  \let\sc=\sevenrm \let\big=\ninebig \normalbaselines\rm}
\gdef\eightpoint{\def\rm{\fam0\eightrm}
  \textfont0=\eightrm \scriptfont0=\sixrm \scriptscriptfont0=\fiverm
  \textfont1=\eighti \scriptfont1=\sixi \scriptscriptfont1=\fivei
  \textfont2=\eightsy \scriptfont2=\sixsy \scriptscriptfont2=\fivesy
  \textfont3=\tenex \scriptfont3=\tenex \scriptscriptfont3=\tenex
  \def\mcal{\fam2 \eightsy}  \def\mmit{\fam1 \eighti}
  \textfont\itfam=\eightit \def\it{\fam\itfam\eightit}
  \textfont\slfam=\eightsl \def\sl{\fam\slfam\eightsl}
  \textfont\ttfam=\eighttt \scriptfont\ttfam=\eighttt
  \scriptscriptfont\ttfam=\eighttt \def\tt{\fam\ttfam\eighttt}
  \textfont\bffam=\eightbf \scriptfont\bffam=\sixbf
  \scriptscriptfont\bffam=\fivebf \def\bf{\fam\bffam\eightbf}
     \ifx\arisposta\amsrisposta   \ifnum\contaeuler=1
  \textfont\eufmfam=\eighteufm \scriptfont\eufmfam=\sixeufm
  \scriptscriptfont\eufmfam=\fiveeufm \def\eufm{\fam\eufmfam\eighteufm}
  \textfont\eufbfam=\eighteufb \scriptfont\eufbfam=\sixeufb
  \scriptscriptfont\eufbfam=\fiveeufb \def\eufb{\fam\eufbfam\eighteufb}
  \def\eurm{\eighteurm} \def\eurb{\eighteurb} \def\eusm{\eighteusm}
  \def\eusb{\eighteusb}       \fi    \ifnum\contaams=1
  \textfont\msamfam=\eightmsam \scriptfont\msamfam=\sixmsam
  \scriptscriptfont\msamfam=\fivemsam \def\msam{\fam\msamfam\eightmsam}
  \textfont\msbmfam=\eightmsbm \scriptfont\msbmfam=\sixmsbm
  \scriptscriptfont\msbmfam=\fivemsbm \def\msbm{\fam\msbmfam\eightmsbm}
     \fi       \ifnum\contacyrill=1     \def\cyrill{\eightwncyr}
  \def\cyrilb{\eightwncyb}  \def\cyrili{\eightwncyi}         \fi
  \textfont3=\eightex \scriptfont3=\sevenex \scriptscriptfont3=\sevenex
  \def\cmmib{\fam\cmmibfam\eightcmmib}  \textfont\cmmibfam=\eightcmmib
  \scriptfont\cmmibfam=\sixcmmib \scriptscriptfont\cmmibfam=\fivecmmib
  \def\cmbsy{\fam\cmbsyfam\eightcmbsy}  \textfont\cmbsyfam=\eightcmbsy
  \scriptfont\cmbsyfam=\sixcmbsy \scriptscriptfont\cmbsyfam=\fivecmbsy
  \def\cmcsc{\fam\cmcscfam\eightcmcsc} \scriptfont\cmcscfam=\eightcmcsc
  \textfont\cmcscfam=\eightcmcsc \scriptscriptfont\cmcscfam=\eightcmcsc
     \fi             \tt \ttglue=.5em plus.25em minus.15em
  \normalbaselineskip=9pt
  \setbox\strutbox=\hbox{\vrule height7pt depth2pt width0pt}
  \let\sc=\sixrm \let\big=\eightbig \normalbaselines\rm }
\gdef\tenbig#1{{\hbox{$\left#1\vbox to8.5pt{}\right.\n@space$}}}
\gdef\ninebig#1{{\hbox{$\textfont0=\tenrm\textfont2=\tensy
   \left#1\vbox to7.25pt{}\right.\n@space$}}}
\gdef\eightbig#1{{\hbox{$\textfont0=\ninerm\textfont2=\ninesy
   \left#1\vbox to6.5pt{}\right.\n@space$}}}
\def\alternativefont#1#2{\ifx\arisposta\amsrisposta \relax \else
\xdef#1{#2} \fi}
\global\contaeuler=0 \global\contacyrill=0 \global\contaams=0
%
%
%
%
\newbox\fotlinebb \newbox\hedlinebb \newbox\leftcolumn
\gdef\makeheadline{\vbox to 0pt{\vskip-22.5pt
     \fullline{\vbox to8.5pt{}\the\headline}\vss}\nointerlineskip}
\gdef\makehedlinebb{\vbox to 0pt{\vskip-22.5pt
     \fullline{\vbox to8.5pt{}\copy\hedlinebb\hfil
     \line{\hfill\the\headline\hfill}}\vss} \nointerlineskip}
\gdef\makefootline{\baselineskip=24pt \fullline{\the\footline}}
\gdef\makefotlinebb{\baselineskip=24pt
    \fullline{\copy\fotlinebb\hfil\line{\hfill\the\footline\hfill}}}
\gdef\doubleformat{\shipout\vbox{\Landspec\makehedlinebb
     \fullline{\box\leftcolumn\hfil\columnbox}\makefotlinebb}
     \advancepageno}
\gdef\columnbox{\leftline{\pagebody}}
\gdef\line#1{\hbox to\hsize{\hskip\leftskip#1\hskip\rightskip}}
\gdef\fullline#1{\hbox to\fullhsize{\hskip\leftskip{#1}%
\hskip\rightskip}}
\gdef\footnote#1{\let\@sf=\empty
         \ifhmode\edef\#sf{\spacefactor=\the\spacefactor}\/\fi
         #1\@sf\vfootnote{#1}}
\gdef\vfootnote#1{\insert\footins\bgroup
         \ifnum\dimnota=1  \eightpoint\fi
         \ifnum\dimnota=2  \ninepoint\fi
         \ifnum\dimnota=0  \tenpoint\fi
         \interlinepenalty=\interfootnotelinepenalty
         \splittopskip=\ht\strutbox
         \splitmaxdepth=\dp\strutbox \floatingpenalty=20000
         \leftskip=\oldssposta \rightskip=\olddsposta
         \spaceskip=0pt \xspaceskip=0pt
         \ifnum\sinnota=0   \textindent{#1}\fi
         \ifnum\sinnota=1   \item{#1}\fi
         \footstrut\futurelet\next\fo@t}
\gdef\fo@t{\ifcat\bgroup\noexpand\next \let\next\f@@t
             \else\let\next\f@t\fi \next}
\gdef\f@@t{\bgroup\aftergroup\@foot\let\next}
\gdef\f@t#1{#1\@foot} \gdef\@foot{\strut\egroup}
\gdef\footstrut{\vbox to\splittopskip{}}
\skip\footins=\bigskipamount
\count\footins=1000  \dimen\footins=8in
\catcode`@=12
\tenpoint
\ifnum\unoduecol=1 \hsize=\tothsize   \fullhsize=\tothsize \fi
\ifnum\unoduecol=2 \hsize=\collhsize  \fullhsize=\tothsize \fi
\global\let\lrcol=L      \ifnum\unoduecol=1
\output{\plainoutput{\ifnum\tipbnota=2 \clearnmbnota\fi}} \fi
\ifnum\unoduecol=2 \output{\if L\lrcol
     \global\setbox\leftcolumn=\columnbox
     \global\setbox\fotlinebb=\line{\hfill\the\footline\hfill}
     \global\setbox\hedlinebb=\line{\hfill\the\headline\hfill}
     \advancepageno  \global\let\lrcol=R
     \else  \doubleformat \global\let\lrcol=L \fi
     \ifnum\outputpenalty>-20000 \else\dosupereject\fi
     \ifnum\tipbnota=2\clearnmbnota\fi }\fi
\def\ifdoublepage{\ifnum\unoduecol=2 }
\gdef\yespagenumbers{\footline={\hss\tenrm\folio\hss}}
\gdef\ciao{ \ifnum\fdefcontre=1 \endfdef\fi
     \par\vfill\supereject \ifnum\unoduecol=2
     \if R\lrcol  \headline={}\nopagenumbers\null\vfill\eject
     \fi\fi \end}

\newskip\olddsposta \newskip\oldssposta
\global\oldssposta=\leftskip \global\olddsposta=\rightskip

\def\filldots{\leaders\hbox to 1em{\hss.\hss}\hfill}
\def\inquadrb#1 {\vbox {\hrule  \hbox{\vrule \vbox {\vskip .2cm
    \hbox {\ #1\ } \vskip .2cm } \vrule  }  \hrule} }
 \def\newline{\hfil\break}
\def\jump{\vskip\baselineskip} \newskip\iinnffrr
\def\sjump{\iinnffrr=\baselineskip
          \divide\iinnffrr by 2 \vskip\iinnffrr}
\def\bjump{\vskip\baselineskip \vskip\baselineskip}
\newcount\nmbnota  \def\clearnmbnota{\global\nmbnota=0}
\newcount\tipbnota \def\letterfootnote{\global\tipbnota=1}

\def\note#1{\global\advance\nmbnota by 1 \ifnum\tipbnota=1
    \footnote{$^{\rm\nttlett}$}{#1} \else {\ifnum\tipbnota=2
    \footnote{$^{\nttsymb}$}{#1}
    \else\footnote{$^{\the\nmbnota}$}{#1}\fi}\fi}
\def\nttlett{\ifcase\nmbnota \or a\or b\or c\or d\or e\or f\or
g\or h\or i\or j\or k\or l\or m\or n\or o\or p\or q\or r\or
s\or t\or u\or v\or w\or y\or x\or z\fi}
\def\nttsymb{\ifcase\nmbnota \or\dag\or\sharp\or\ddag\or\star\or
\natural\or\flat\or\clubsuit\or\diamondsuit\or\heartsuit
\or\spadesuit\fi}   \clearnmbnota
\def\numberfootnote{\global\tipbnota=0} \numberfootnote
\def\setnote#1{\expandafter\xdef\csname#1\endcsname{
\ifnum\tipbnota=1 {\rm\nttlett} \else {\ifnum\tipbnota=2
{\nttsymb} \else \the\nmbnota\fi}\fi} }
\newcount\nbmfig  \def\clearnbmfig{\global\nbmfig=0}
\gdef\figure{\global\advance\nbmfig by 1
      {\rm fig. \the\nbmfig}}   \clearnbmfig
\def\setfig#1{\expandafter\xdef\csname#1\endcsname{fig. \the\nbmfig}}
 \def\endformula{\eqno\numero $$}
 \def\efr{\endformula}
\newcount\frmcount \def\clearfrmcount{\global\frmcount=0}
\def\numero{\global\advance\frmcount by 1   \ifnum\indappcount=0
  {\ifnum\cpcount <1 {\hbox{\rm (\the\frmcount )}}  \else
  {\hbox{\rm (\the\cpcount .\the\frmcount )}} \fi}  \else
  {\hbox{\rm (\applett .\the\frmcount )}} \fi}
\def\nameformula#1{\global\advance\frmcount by 1%
\ifnum\draftnum=0  {\ifnum\indappcount=0%
{\ifnum\cpcount<1\xdef\spzzttrra{(\the\frmcount )}%
\else\xdef\spzzttrra{(\the\cpcount .\the\frmcount )}\fi}%
\else\xdef\spzzttrra{(\applett .\the\frmcount )}\fi}%
\else\xdef\spzzttrra{(#1)}\fi%
\expandafter\xdef\csname#1\endcsname{\spzzttrra}
\eqno \hbox{\rm\spzzttrra} $$}
\def\nfr{\nameformula}    
\def\nameali#1{\global\advance\frmcount by 1%
\ifnum\draftnum=0  {\ifnum\indappcount=0%
{\ifnum\cpcount<1\xdef\spzzttrra{(\the\frmcount )}%
\else\xdef\spzzttrra{(\the\cpcount .\the\frmcount )}\fi}%
\else\xdef\spzzttrra{(\applett .\the\frmcount )}\fi}%
\else\xdef\spzzttrra{(#1)}\fi%
\expandafter\xdef\csname#1\endcsname{\spzzttrra}
  \hbox{\rm\spzzttrra} }      \clearfrmcount
\newcount\cpcount \def\clearcpcount{\global\cpcount=0}
\newcount\subcpcount \def\clearsubcpcount{\global\subcpcount=0}
\newcount\appcount \def\clearappcount{\global\appcount=0}
\newcount\indappcount \def\clearindappcount{\indappcount=0}
\newcount\sottoparcount 

\def\applett{\ifcase\appcount  \or {A}\or {B}\or {C}\or
{D}\or {E}\or {F}\or {G}\or {H}\or {I}\or {J}\or {K}\or {L}\or
{M}\or {N}\or {O}\or {P}\or {Q}\or {R}\or {S}\or {T}\or {U}\or
{V}\or {W}\or {X}\or {Y}\or {Z}\fi    \ifnum\appcount<0
\immediate\write16 {Panda ERROR - Appendix: counter "appcount"
out of range}\fi  \ifnum\appcount>26  \immediate\write16 {Panda
ERROR - Appendix: counter "appcount" out of range}\fi}
\clearappcount  \clearindappcount \newcount\connttrre
\def\clearconnttrre{\global\connttrre=0} \newcount\countref
\def\clearcountref{\global\countref=0} \clearcountref
\def\chapter#1{\global\advance\cpcount by 1 \clearfrmcount
                 \goodbreak\null\vbox{\jump\nobreak
                 \clearsubcpcount\clearindappcount
                 \itemitem{\ttaarr\the\cpcount .\qquad}{\ttaarr #1}
                 \par\nobreak\jump\sjump}\nobreak}
\def\section#1{\global\advance\subcpcount by 1 \goodbreak\null
               \vbox{\sjump\nobreak\ifnum\indappcount=0
                 {\ifnum\cpcount=0 {\itemitem{\ppaarr
               .\the\subcpcount\quad\enskip\ }{\ppaarr #1}\par} \else
                 {\itemitem{\ppaarr\the\cpcount .\the\subcpcount\quad
                  \enskip\ }{\ppaarr #1} \par}  \fi}
                \else{\itemitem{\ppaarr\applett .\the\subcpcount\quad
                 \enskip\ }{\ppaarr #1}\par}\fi\nobreak\jump}\nobreak}
\clearsubcpcount
\def\appendix#1{\global\advance\appcount by 1 \clearfrmcount
                  \goodbreak\null\vbox{\jump\nobreak
                  \global\advance\indappcount by 1 \clearsubcpcount
          \itemitem{ }{\hskip-40pt\ttaarr Appendix\ \applett :\ #1}
             \nobreak\jump\sjump}\nobreak}
\clearappcount \clearindappcount
\def\references{\goodbreak\null\vbox{\jump\nobreak
   \itemitem{}{\ttaarr References} \nobreak\jump\sjump}\nobreak}

\clearcpcount\clearcountref

\def\setchap#1{\ifnum\indappcount=0{\ifnum\subcpcount=0%
\xdef\spzzttrra{\the\cpcount}%
\else\xdef\spzzttrra{\the\cpcount .\the\subcpcount}\fi}
\else{\ifnum\subcpcount=0 \xdef\spzzttrra{\applett}%
\else\xdef\spzzttrra{\applett .\the\subcpcount}\fi}\fi
\expandafter\xdef\csname#1\endcsname{\spzzttrra}}
\newcount\draftnum \newcount\ppora   \newcount\ppminuti
\global\ppora=\time   \global\ppminuti=\time
\global\divide\ppora by 60  \draftnum=\ppora
\multiply\draftnum by 60    \global\advance\ppminuti by -\draftnum
\def\droggi{\number\day /\number\month /\number\year\ \the\ppora
:\the\ppminuti}     \global\draftnum=0
\def\draftcomment#1{\ifnum\draftnum=0 \relax \else
{\ {\bf ***}\ #1\ {\bf ***}\ }\fi} 
%
%
\catcode`@=11
\gdef\Ref#1{\expandafter\ifx\csname @rrxx@#1\endcsname\relax%
{\global\advance\countref by 1    \ifnum\countref>200
\immediate\write16 {Panda ERROR - Ref: maximum number of references
exceeded}  \expandafter\xdef\csname @rrxx@#1\endcsname{0}\else
\expandafter\xdef\csname @rrxx@#1\endcsname{\the\countref}\fi}\fi
\ifnum\draftnum=0 \csname @rrxx@#1\endcsname \else#1\fi}
\gdef\beginref{\ifnum\draftnum=0  \gdef\Rref{\fairef}
\gdef\endref{\scriviref} \else\relax\fi
\ifx\risposta\mplarisposta \ninepoint \fi
\parskip 2pt plus.2pt \baselineskip=12pt}
\def\Reflab#1{[#1]} \gdef\Rref#1#2{\item{\Reflab{#1}}{#2}}
\gdef\endref{\relax}  \newcount\conttemp
\gdef\fairef#1#2{\expandafter\ifx\csname @rrxx@#1\endcsname\relax
{\global\conttemp=0 \immediate\write16 {Panda ERROR - Ref: reference
[#1] undefined}} \else
{\global\conttemp=\csname @rrxx@#1\endcsname } \fi
\global\advance\conttemp by 50  \global\setbox\conttemp=\hbox{#2} }
\gdef\scriviref{\clearconnttrre\conttemp=50
\loop\ifnum\connttrre<\countref \advance\conttemp by 1
\advance\connttrre by 1
\item{\Reflab{\the\connttrre}}{\unhcopy\conttemp} \repeat}
\clearcountref \clearconnttrre
\catcode`@=12
\ifx\risposta\mplarisposta \def\Reflab#1{#1.} \letterfootnote \fi

\def\slashchar#1{\setbox0=\hbox{$#1$} \dimen0=\wd0
     \setbox1=\hbox{/} \dimen1=\wd1 \ifdim\dimen0>\dimen1
      \rlap{\hbox to \dimen0{\hfil/\hfil}} #1 \else
      \rlap{\hbox to \dimen1{\hfil$#1$\hfil}} / \fi}
\ifx\oldchi\undefined \let\oldchi=\chi
  \def\cchi{{\raise 1pt\hbox{$\oldchi$}}} \let\chi=\cchi \fi
\def\square{\hbox{{$\sqcup$}\llap{$\sqcap$}}}

\def\frac#1#2{{\textstyle{#1 \over #2}}}

\def\half{\ifinner {\scriptstyle {1 \over 2}}\else {1 \over 2} \fi}

\def\simge{\rlap{\raise 2pt \hbox{$>$}}{\lower 2pt \hbox{$\sim$}}}
\def\simle{\rlap{\raise 2pt \hbox{$<$}}{\lower 2pt \hbox{$\sim$}}}

\def\vbig#1#2{{\vbigd@men=#2\divide\vbigd@men by 2%
\hbox{$\left#1\vbox to \vbigd@men{}\right.\n@space$}}}

%
%
\newcount\fdefcontre \newcount\fdefcount \newcount\indcount
\newread\filefdef  \newread\fileftmp  \newwrite\filefdef
\newwrite\fileftmp     \def\strip#1*.A {#1}
\def\futuredef#1{\beginfdef
\expandafter\ifx\csname#1\endcsname\relax%
{\immediate\write\fileftmp {#1*.A}
\immediate\write16 {Panda Warning - fdef: macro "#1" on page
\the\pageno \space undefined}
\ifnum\draftnum=0 \expandafter\xdef\csname#1\endcsname{(?)}
\else \expandafter\xdef\csname#1\endcsname{(#1)} \fi
\global\advance\fdefcount by 1}\fi   \csname#1\endcsname}

\def\beginfdef{\ifnum\fdefcontre=0
\immediate\openin\filefdef \jobname.fdef
\immediate\openout\fileftmp \jobname.ftmp
\global\fdefcontre=1  \ifeof\filefdef \immediate\write16 {Panda
WARNING - fdef: file \jobname.fdef not found, run TeX again}
\else \immediate\read\filefdef to\spzzttrra
\global\advance\fdefcount by \spzzttrra
\indcount=0      \loop\ifnum\indcount<\fdefcount
\advance\indcount by 1   \immediate\read\filefdef to\spezttrra
\immediate\read\filefdef to\sppzttrra
\edef\spzzttrra{\expandafter\strip\spezttrra}
\immediate\write\fileftmp {\spzzttrra *.A}
\expandafter\xdef\csname\spzzttrra\endcsname{\sppzttrra}
\repeat \fi \immediate\closein\filefdef \fi}
\def\endfdef{\immediate\closeout\fileftmp   \ifnum\fdefcount>0
\immediate\openin\fileftmp \jobname.ftmp
\immediate\openout\filefdef \jobname.fdef
\immediate\write\filefdef {\the\fdefcount}   \indcount=0
\loop\ifnum\indcount<\fdefcount    \advance\indcount by 1
\immediate\read\fileftmp to\spezttrra
\edef\spzzttrra{\expandafter\strip\spezttrra}
\immediate\write\filefdef{\spzzttrra *.A}
\edef\spezttrra{\string{\csname\spzzttrra\endcsname\string}}
\iwritel\filefdef{\spezttrra}
\repeat  \immediate\closein\fileftmp \immediate\closeout\filefdef
\immediate\write16 {Panda Warning - fdef: Label(s) may have changed,
re-run TeX to get them right}\fi}
\def\iwritel#1#2{\newlinechar=-1
{\newlinechar=`\ \immediate\write#1{#2}}\newlinechar=-1}
\global\fdefcontre=0 \global\fdefcount=0 \global\indcount=0
%
%
\null
%
%
%

\loadamsmath
\nopagenumbers{\baselineskip=12pt
\line{\hfill OUTP-92-19P}
\line{\hfill September, 1992}
\line{\hfill hep-th/9209024}
\ifdoublepage \bjump\bjump\bjump\bjump\else\vfill\fi
\centerline{\capsone QUANTUM SOLITON MASS CORRECTIONS IN}
\sjump
\centerline{\capsone SL(N) AFFINE TODA FIELD THEORY}
\bjump\bjump
\centerline{\tencmcsc Timothy Hollowood\footnote{$^*$}
{Email: holl\%dionysos.thphys@prg.oxford.ac.uk\newline
Address after $1^{\rm st}$ October 1992: Theory Division,
CERN, 1211 Geneva 23, Switzerland}}
\sjump
\centerline{\sl Theoretical Physics, 1 Keble Road}
\centerline{\sl Oxford, OX1 3NP, U.K.}
\vfill
\ifnum\unoduecol=2 \eject\null\vfill\fi
\centerline{\capsone ABSTRACT}
\sjump
\noindent The first quantum mass corrections for the solitons of
complex $sl(n)$ affine Toda field theory are calculated. The
corrections are real and
preserve the classical mass ratios. The formalism also proves that the
solitons are classically stable.
\bjump\bjump
\ifnum\unoduecol=2 \vfill\fi
\eject}
\yespagenumbers\pageno=1

\chapter{Introduction}

The equation of motion for the complex $sl(n)^{(1)}$ Toda field theory
can be written
$$
\square\phi=-{m^2\over i\beta}
\sum_{j=1}^n\alpha_j\exp\,i\beta\alpha_j\cdot\phi.
\nfr{EQMOT}
The field $\phi(x,t)$ is an $n-1$-dimensional vector: an element of the
Cartan subalgebra of the finite Lie algebra $sl(n)$.
The inner products are taken
with respect to the Killing form of $sl(n)$ restricted to the Cartan
subalgebra. The $\alpha_j$'s, for $j=1,2,\ldots,n-1$ are the simple roots
of $sl(n)$; $\alpha_0$ is the extended root (minus the highest root of
$sl(n)$). The fact that the extended root is included in the sum,
distinguishes the {\it affine\/} theories from the {\it non-affine\/} ones.
The $\alpha_j$'s are linearly dependent:
$$
\sum_{j=1}^n\alpha_j=0.
\efr
The constant $m$ sets an arbitrary mass scale and $\beta$ is a
coupling constant. In the {\it complex\/} theories $\beta$ is real,
whereas, in the {\it real\/} theories $\beta$ is purely imaginary.

The {\it real\/} theories, when $\tilde\beta=i\beta\in{\Bbb R}$, are well
understood, both in the classical and quantum regime (see for example
[\Ref{todasm}]). The spectrum consists of $n-1$ particles of mass
$$
m_a=2m\sin{\pi a\over n},\qquad a=1,2\ldots,n-1.
\nfr{FUNDMASS}
In the quantum theory the spectrum is preserved, except for an
overall renormalization of the mass scale $m$. It is known from a
Feynman diagram calculation to one-loop [\Ref{todasm}], that
$$
\hat m_a=m_a\left[1-{\tilde\beta^2\over 4n}{\rm cot}{\pi\over n}+
O(\tilde\beta^4)\right].
\nfr{QUANMASS}
(In the following we shall use ``hats'' to denote exact quantum masses.)
The {\it complex\/} theories, when $\beta\in{\Bbb R}$, have a much
more complicated spectrum, since they admits kink or soliton
solutions. This is well-known when $n=2$, for which the
real case is the sinh-Gordon theory and the complex case is the
sine-Gordon theory. Soliton solutions were first written
down in ref. [\Ref{SOL}] for the affine $sl(n)$ and $d_4$
theories; recently, soliton solutions have been found for all the
affine Toda theories [\Ref{GENSOL}].

For the $sl(n)$ theories the general $N$-soliton solution is
constructed in the following way. To each soliton one associates the
data $\{\sigma,\lambda,a,\xi\}$, where $\sigma$ and $\lambda$ are real
parameters satisfying
$$
{\cal F}(\sigma,\lambda,a)=0,
\efr
where the characteristic polynomial is
$$
{\cal F}(\sigma,\lambda,a)=\sigma^2-\lambda^2-4m^2\sin^2{\pi
a\over n},
\efr
$a$ is an integer in the set $\{1,2,\ldots,n-1\}$ and $\xi$ is, for
the moment, an arbitrary complex parameter. To each soliton, say the
$p^{\rm th}$, one associates the $n$ functions
$$
\Phi_j^{(p)}(x,t)=\sigma_px-\lambda_pt+{2\pi i\over
n}a_pj+\xi_p,\qquad j=1,2,\ldots,n,
\efr
and to each soliton pair, say the $p^{\rm th}$ and $q^{\rm th}$, one
associates the ``interaction function''
$$
\exp\gamma^{(pq)}=-{{\cal
F}(\sigma_p-\sigma_q,\lambda_p-\lambda_q,a_p-a_q)
\over{\cal F}(\sigma_p+\sigma_q,\lambda_p+\lambda_q,a_p+a_q)}.
\nfr{INT}
The general $N$-soliton solution to the equations of motion is
$$
\phi=-{1\over i\beta}\sum_{j=1}^n\alpha_j\log\left[\sum_{\mu_1=0}^1
\cdots\sum_{\mu_N=0}^1\exp\left(\sum_{p=1}^N\mu_p\Phi_j^{(p)}+
\sum_{1\leq p<q\leq N}\mu_p\mu_q\gamma^{(pq)}\right)\right].
\nfr{NSOL}
So the one-soliton solution is
$$
\phi(x,t)=-{1\over i\beta}\sum_{j=1}^n\alpha_j\log\left[1+\exp
\left(\sigma x-\lambda t+\xi+{2\pi ia\over n}j\right)\right],
\nfr{ONESOL}
with
$$
\sigma^2-\lambda^2=4m^2\sin^2{\pi a\over n}.
\efr
The soliton is a kink, whose centre-of-mass is at
$\sigma^{-1}[\lambda t-{\rm Re}\,\xi]$, moving with velocity
$\lambda/\sigma$ and having characteristic size $\sigma^{-1}$. The
parameters $a$ and ${\rm Im}\,\xi$ determine the topological charge of
the soliton. The topological charge is defined to be
$$
t={\beta\over2\pi}\int_{-\infty}^\infty dx\,{\partial\phi\over\partial
x},
\nfr{TOP}
from which one readily verifies that \ONESOL\ has a topological charge
which a weight of the $a^{\rm th}$ fundamental representation of
$sl(n)$, where the exact weight is determined by ${\rm Im}\,\xi$.

The classical masses of the solitons are easily determined by explicit
calculation from the Hamiltonian:
$$
H={1\over2}\left({\partial\phi\over\partial
t}\right)^2+{1\over2}\left({\partial\phi\over\partial
x}\right)^2-{m^2\over\beta^2}\sum_{j=1}^n\left[\exp(i\beta\alpha_j\cdot\phi)-1
\right].
\nfr{HAM}
One finds
$$
M_a={4mn\over\beta^2}\sin{\pi a\over n},\qquad
a=1,2,\ldots,n-1.
\nfr{CSOLMASS}
Notice that these masses are real, despite the complex form for the
Hamiltonian. More particularly, notice that they
are proportional to the fundamental masses of \FUNDMASS: $M_a\propto
m_a$. Both these facts were explained in [\Ref{SOL}].

In ref. [\Ref{SM}] the quantum theory of the solitons
was investigated. A form for the soliton-soliton $S$-matrix was proposed, which
agrees with the WKB quantization of the classical scattering theory,
discussed above, in the semi-classical limit. In general, it was shown
that the $S$-matrix represents a non-unitary quantum field theory,
but that there was a unitary region depending on the coupling
constant, for which, in addition, the bootstrap closes on a finite set
of states corresponding to solitons transforming in the fundamental
representations of $sl(n)$. Central to the proposal in ref. [\Ref{SM}]
for the quantum theory of the solitons, is the idea that
the ratios of the soliton masses should not be altered in the quantum
theory: $\hat M_a\propto m_a$. It is clearly of interest to
calculate the quantum corrections to the soliton masses directly,
in order to verify the truth of this fact.

There exists a standard method for computing the first quantum
corrections to soliton masses in a $1+1$-dimensional field theory: one
basically sums the zero-point energies of the fluctuations around the
soliton solution. Obviously, such a sum is divergent, but these
divergences may be removed, in the standard way, by renormalization.
Moreover, there is a piece of folklore for
$1+1$-integrable theories that says that the masses of solitons
obtained in this way are exact and there are no higher-order
corrections. This is appears to be true for the sine-Gordon theory (the
case for $g=sl(2)$), where the exact soliton mass is
$$
\hat M={8m\over\beta^2}-{2m\over\pi},
\efr
in our conventions --- the first term being the classical contribution
from \CSOLMASS, and the second term being the first quantum
correction. This piece of folklore can, to some extent, be justified by noting
that at some level the theory can be cast, via a transformation to
action-angle variables, into a set of harmonic oscillators
--- albeit of infinite number --- and the
saddle-point calculation of the soliton masses should, therefore, be exact.
Although this argument has a certain verisimilitude, it is not clear
how general the principle is; in particular, does it apply to Toda
theories, beyond the sine-Gordon theory?

As a by-product of our calculation of the quantum corrections to the
soliton masses, we are able to prove that the soliton solutions are
classically stable; an issue that is important given the fact that the
Hamiltonian \HAM\ is complex.

\chapter{Classical Stability}

\def\pb{\bar\phi}

In this section, we consider the question of the classical stability
of the one-soliton solutions. This question is closely related to
the calculation of the quantum corrections to the soliton masses, as
we shall see.

\def\et{\tilde\eta}

The standard way to approach the question of stability, is to consider
the effect of adding a small perturbation $\eta$ to the one-soliton
solution \ONESOL, which we denote $\pb$. It is convenient to take the
soliton to be stationary with the centre-of-mass at the origin:
$\lambda=0$, $\sigma=m_a$ and ${\rm Re}\,\xi=0$. To first order
in $\eta$ the equation of motion \EQMOT\ becomes
$$
\square\eta+m^2\sum_{j=1}^n\alpha_j(\alpha_j\cdot\eta)\exp
\,i\beta\alpha_j\cdot\pb=0.
\nfr{perteq}
Now consider a perturbation with a time-dependence of the form
$\eta(x,t)=\tilde\eta(x)\exp\,i\nu t$, so that
$$
{\cal D}\,\et=\nu^2\et,
\nfr{SCH}
where we have defined the following differential operator:
$$
{\cal D}=-{\partial^2\over\partial x^2}+m^2\sum_{j=1}^n\alpha_j\otimes\alpha_j
\,\exp\,i\beta\alpha_j\cdot\pb(x).
\nfr{DIFOP}
The question of stability now boils down to showing that the spectrum
of ${\cal D}$ --- for bounded eigenfunctions ---
is real and positive; hence, the frequencies $\nu$ are real and
small perturbations to $\pb$ do not diverge. Our strategy for proving
this will be to find the exact spectrum of ${\cal D}$.

First of all, the spectrum of ${\cal D}$
is real, since it is actually a
hermitian operator, with respect to some inner product.
To show this, we first notice that the one-soliton solutions --- for
some particular choice of ${\rm Im}\,\xi$ --- actually satisfy a
reality condition of the form
$$
\pb^\star(x)=-M\pb(x),
\nfr{CC}
where $M$ acts as a ${\Bbb Z}_2$ symmetry of the roots $\alpha_j$:
$$
M\alpha_j=\alpha_{m(j)},\qquad m^2(j)=j.
\efr
This is explained in detail in refs. [\Ref{SOL},\Ref{JON}]. From this
it follows that
$$
{\cal D}^\dagger=M{\cal D}M.
\efr
Hence, ${\cal D}$ is hermitian with respect to the inner product
$$
(f,g)=\int_{-\infty}^\infty dx\,f^\dagger\cdot Mg,
\efr
for functions $f(x)$ and $g(x)$ in the Cartan subalgebra of $sl(n)$.

We now turn to the determination of the spectrum of bounded
eigenfunctions of ${\cal D}$.
The eigenvalue equation \SCH\ can be viewed as a multi-component
Schr\"odinger problem for the potential
$$
V(x)=\sum_{j=1}^n\alpha_j\otimes\alpha_j\,\exp\,i\beta\alpha_j\cdot\pb.
\nfr{POT}
The spectrum has contributions from two sources: the bound states
and the scattering solutions. The former have a discrete
spectrum and the latter a continuous spectrum. There are also
eigenfunctions which are not bounded, but these have no relevance for
the discussion of stability and for the quantum mass corrections.

In general the scattering solutions have the following asymptotic behaviour
$$\eqalign{
\lim_{x\rightarrow-\infty}\et(x)&=A\exp\,ikx+B(k)\exp\,-ikx\cr
\lim_{x\rightarrow\infty}\et(x)&=C(k)\exp\,ikx,\cr}
\nfr{ASY}
where the incoming, reflection and transmission coefficients, $A$,
$B(k)$ and $C(k)$, are Cartan subalgebra-valued.
Remarkably, the potential in \POT\ is
``reflection-less'', so that $B(k)=0$. This can be shown directly
using the following argument. In the introduction, we wrote
down an expression for the $N$-soliton solution. Consider a
two-soliton solution where the first soliton is $\pb$,
the static one-soliton solution that we started with, and the second
soliton --- the ``probe'' --- will act as a small perturbation of the
first. To this end, we take $\sigma_2=ik$, $a_2=b$,
$\lambda_2=-i\nu$ and treat $\exp\,\xi_2$ as
a small parameter; in which case to first order in $\exp\,\xi_2$ we deduce
$$
\et_b(k;x)=
\sum_{j=1}^n\alpha_j\left[{1+\omega^{aj}\exp(m_a
x+\xi+\gamma)\over1+\omega^{aj}\exp(m_ax+\xi)}\right]\omega^{bj}
\exp\,ikx,
\nfr{fullsol}
where we have introduced $\omega$, the primitive
$n^{\rm th}$ root of unity. In the above, $\gamma$ is the interaction
parameter \INT\ which gives in this case
$$
\exp\,\gamma(k)=-{m_a^2+m_b^2-m_{a-b}^2-2im_a k\over m_a^2+m_b^2-m_{a+b}^2
+2im_ak}.
\nfr{intpar}
The expression in \fullsol\ is the exact scattering solution to the
Schr\"odinger problem \SCH\ with eigenvalue $\nu^2=k^2+m_b^2$ and
$$
B(k)=0,\qquad A=\sum_{j=1}^n\omega^{bj}\alpha_j,\qquad
C(k)=\exp\,\gamma(k)\sum_{j=1}^n\omega^{bj}\alpha_j.
\nfr{ASYCOND}
Moreover, the set of solutions $\et_b(k;x)$, for $b=1,2,\ldots,n-1$ and
$k\in{\Bbb R}$, forms a complete set of scattering solutions, since the
vectors $\sum_{j=1}^n\omega^{bj}\alpha_j$, for $b=1,2,\ldots,n-1$, span
the Cartan subalgebra. Hence, \POT\ is a reflection-less potential.

The beauty of dealing with a reflection-less potential is that it
allows for a simple
determination of the bound-states as well. These solutions occur for
values of $k$ for which the transmission coefficient has a zero, and
which are, in addition, normalizable. It is
easy to see, from the explicit expression for $\exp\gamma(k)$ in \intpar,
that $C(k)$ has a zero when
$$
k={m_a^2+m_b^2-m_{a-b}^2\over 2im_a}=-im_b\cos{\pi(a-b)\over n}.
\nfr{KVAL}
The exact solution for the bound-states
follows from \fullsol\ and the particular value for $k$ in \KVAL:
$$
\et_b(x)=\sum_{j=1}^n\alpha_j{\omega^{bj}\exp\left[m_bx\cos{\pi(a-b)
\over n}\right]\over1+\omega^{aj}\exp(m_ax+\xi)},
\nfr{BS}
which has a frequency
$$
\nu=m_b\left\vert\sin{\pi(a-b)\over n}\right\vert.
\nfr{FREQBS}
The asymptotic limits of the above solution are
$$
\eqalign{
\lim_{x\rightarrow-\infty}\et(x)&=\sum_{j=1}^n\alpha_j\omega^{bj}
\exp\left[2mx\sin{\pi b\over n}\cos{\pi(a-b)
\over n}\right]\cr
\lim_{x\rightarrow\infty}\et(x)&=\sum_{j=1}^n\alpha_j\omega^{(b-a)j}
\exp\left[2mx\sin{\pi(a-b)\over n}\cos{\pi
b\over n}-\xi\right].\cr}
\nfr{ASYBS}
Of course, not all these solutions have the right asymptotics to be
{\it bona-fide\/} bound-states of the Schr\"odinger problem. The
{\it bone-fide\/} bound-states must be normalizable which means
$$
\lim_{x\rightarrow\pm\infty}\et(x)\sim\exp\,\mp|\kappa^{\pm}|x,
\efr
for some real non-zero constants $\kappa^\pm\neq0$.
The values of $b$ which lead to {\it bona-fide\/} bound-states are
$$
\eqalign{
&1\leq b<a\quad{\rm or}\quad {n\over2}<b<{n\over2}+a,\qquad {\rm
for}\ a\leq{n\over2}\cr
&a-{n\over2}<b<{n\over2}\quad{\rm or}\quad a<b\leq n-1,\qquad {\rm
for}\ a\geq{n\over2}.\cr}
\nfr{RANG}
The solution with $b=a$ has not been included since this corresponds
to the zero-mode of ${\cal D}$ given by
$$
\tilde\eta_a(x)=\sum_{j=1}^n\alpha_j{\omega^{aj}\exp\,m_ax
\over1+\omega^{aj}\exp\left(m_ax+\xi\right)}.
\nfr{ZM}
The appearance of this zero-mode was only to be expected since
it is proportional to
$\partial\pb/\partial x$, and is due to the freedom to shift the
centre-of-mass of the soliton.

In addition to these bound-states there are
bounded solutions which must also be considered.
Such eigenfunctions have a constant asymptote, and hence are not
normalizable. They appear only when $n$ is even and
are given by \BS\ when
$b=\frac{n}{2}$ or $b=a+\frac{n}{2}$, if $a<\frac{n}{2}$, and
$b=a-\frac{n}{2}$ or $b=\frac{n}{2}$, if $a>\frac{n}{2}$.

{}From the preceeding analysis, we conclude that the spectrum of
bounded eigenfunctions of ${\cal D}$ is real and positive,
except for the zero-mode
which reflects the freedom to move the centre-of-mass of the soliton.
Hence the one-soliton solutions are classically stable to small perturbations.

\chapter{Quantum Corrections to the Masses}

In this section we calculate the first quantum corrections to the
soliton masses. We employ the semi-classical WKB method described in
[\Ref{wkb}] for the kinks, or solitons, of the $\phi^4$ and
sine-Gordon theories --- see also the review in [\Ref{raj}]. The
method is completely standard; however, we feel that the Toda theories
are sufficiently more complicated than these other examples
that it is worthwhile including all the calculational details. In
particular, the question of boundary conditions is not completely standard.

The dimensionless expansion parameter in our theory is
$\hbar\beta^2/m^2$, and so the expansion in $\hbar$ coincides with the
weak-coupling expansion.
{}From now on we set $\hbar=1$. The WKB method of [\Ref{wkb}] is
straightforward to apply in the present situation since the one-soliton
solution is, in its rest-frame,
time-independent. The idea is to compute the zero-point energy of the
small oscillations around the classical solution. The sum over all the
modes can be done when an infra-red regulator is introduced; this is
most easily achieved by putting the theory in a box with rigid
boundary conditions. The
zero-point energy of the vacuum must then be subtracted. The resulting
expression is independent of the size of the box, $L$, as
$L\rightarrow\infty$; however, the final expression has an
ultra-violet divergence. This divergence is not a problem of the
soliton solution {\it per se\/},
rather, it is just a manifestation of the fact that the
bare-mass needs to be renormalized. This is achieved simply by
normal-ordering the Hamiltonian. The extra correction removes the divergence
and the resulting finite residue which remains is then the mass
correction.

The small fluctuations which contribute to the mass shift come from
either the bound-states or the scattering solutions of the linearized
equation of motion around the soliton solution: the Schr\"odinger problem \SCH.

\section{Contributions to the Mass from the Bound-States}

Each of the bound-states $\et_b(x)$, with $b$ in the range \RANG,
contributes an amount $\half\nu$ to the mass shift, where
$\nu$ is given by \FREQBS. In addition to this, the
bounded solutions --- which only occur for $n$ even --- also
contribute to the mass shift. However, we shall not include the states
\BS\ with $b=a+\frac{n}{2}$, for $a<\frac{n}{2}$, and $b=a-\frac{n}{2}$, for
$a>\frac{n}{2}$, which correspond to the $k=0$ solutions of \fullsol,
and whose contribution to the mass shift
will be included in the contribution from the scattering states
discussed in the next subsection. So the contributing modes of the
form \BS\ are
$$
\eqalign{
&1\leq b<a\quad{\rm or}\quad {n\over2}\leq b<{n\over2}+a,\qquad
{\rm for}\ a<{n\over2}\cr
&a-{n\over2}<b\leq{n\over2}\quad{\rm or}\quad a<b\leq n-1,\qquad
{\rm for}\ a>{n\over2}\cr
&1\leq b<{n\over2}\quad{\rm or}\quad{n\over2}<b\leq n-1,\qquad{\rm
for}\ a=\frac{n}{2}.\cr}
\efr
The contribution to the mass shift from these states,
for $a<\frac{n}{2}$, is
$$
\Delta M_a(1)=m\left[\sum_{b=1}^{a-1}\sin{\pi
b\over n}\sin{\pi(a-b)\over
n}-\sum_{b\geq\frac{n}{2}}^{<\frac{n}{2}+a}\sin{\pi b\over
n}\sin{\pi(a-b)\over n}\right].
\efr
(Notice, that the expression is also valid for $a=\frac{n}{2}$, since
in that case the term with $b=\frac{n}{2}$ does not contribute.)
It is straightforward to evaluate the sums involved, giving
$$
\Delta M_a(1)=\cases{{1\over2}m_a\left({\rm
cot}{\pi\over n}+{\rm cosec}{\pi\over
n}\right)\quad&$n\in{\rm Odd}$\cr m_a{\rm cot}{\pi\over n}
&$n\in{\rm Even}$.\cr}
\nfr{BSMC}
It is not difficult to show that the cases when $a>\frac{n}{2}$ give
the same results as \BSMC, as a function of $a$.

\section{Contributions to the Mass from the Scattering States}

In this subsection, we calculate the contribution of the scattering
solutions of \SCH\ to the mass correction. The idea is to sum
the zero-point energies of all the scattering modes. Such a sum will,
of course, be infra-red divergent, so first of all it is
necessary to introduce some finite boundary conditions so that the
modes become discrete and therefore enumerable. Usually,
periodic boundary conditions are chosen; however, they are not appropriate
here because $\gamma(k)$ in \intpar\ is not purely imaginaginary.
It turns out that the appropriate boundary conditions are
$\et(x=0)=\et(x=L)=0$, where the centre-of-mass of the
soliton lies somewhere inside the box, and $L$ is much
larger than the size of the soliton: $L\gg m_a^{-1}$. The solutions
satisfying these boundary conditions are
$$
\et_{b}(k_p;x)-\et_{b}(-k_p;x),
\efr
where $\et_b(k;x)$ is the solution in \fullsol, with
$$
k_pL+\rho_{b}(k_p)=\pi p,\qquad k_p\geq0,\quad p\in{\Bbb Z}.
\nfr{periodic}
where $\rho_{b}(k)={\rm Im}\,\gamma(k)$ and $\gamma(k)$ is defined in
\intpar.

The quantum correction to the soliton mass is then obtained by taking
the zero-point energy of the modes around the soliton solution
and subtracting the zero-point energy of modes around the
vacuum, which satisfy \periodic\ with $\rho=0$. That is
$$
\Delta
M_a(2)=\frac{1}{2}\sum_{b=1}^{n-1}\sum_{k_p\geq0}\left\{\sqrt{k_p^2+
m_b^2}-\sqrt{[k_p+\rho_{b}(k_p)/L]^2+m_b^2}\right\},
\efr
where the sum over $k_p$ is the sum over distinct solutions
of \periodic. Since $L$ will eventually be taken to
infinity, we can expand in $L^{-1}$, the leading term in the sum being
$$
-L^{-1}{k\rho_{b}(k)\over\sqrt{k^2+m_b^2}}=-L^{-1}
\rho_{b}(k){d\epsilon_{b}(k)\over dk},
\efr
with $\epsilon_{b}(k)=\sqrt{k^2+m_b^2}$. Now, we take
$L\rightarrow\infty$ in which case we can replace the sum over $k_p$ by
an integral over $k$:
$$
\sum_{p\geq0}=L\left[\int_{-\infty}^\infty{dk\over2\pi}+O(L^{-1})\right],
\efr
where we have used the fact that the integrand is an even function of $k$.
Therefore, the mass correction is
$$
\eqalign{\Delta
M_a&=-\sum_{b=1}^{n-1}\int_{-\infty}^\infty{dk\over4\pi}\rho_{
b}(k){d\epsilon_{b}(k)\over dk}\cr
&=-{1\over4\pi}\sum_{b=1}^{n-1}\left[\rho_{b}(k)\epsilon_{b}(k)\big
\vert_{-\infty}^\infty+\int_{-\infty}^\infty dk\,\epsilon_{b}(k){
d\rho_{b}(k)\over dk}\right].\cr}
\nfr{twoconts}
Now as $k\rightarrow\pm\infty$, $\epsilon_b(k)\rightarrow|k|$ and
$$
\rho_{b}(k)\rightarrow{1\over2m_ak}\left(2m_a^2+2m_b^2-m_{a-b}^2
-m_{a+b}^2\right)+\pi.
\efr
and thus the first contribution from \twoconts\ is
$$-{1\over4\pi}\sum_{b=1}^{n-1}\rho_{b}(k)\epsilon_{b}(k)
\big\vert_{-\infty}^\infty=-{1\over4\pi
m_a}\sum_{b=1}^{n-1}\left(2m_a^2+2m_b^2-m_{a-b}^2-m_{a+b}^2
\right).
\efr
The sum is straightforward to perform, the result being a contribution
$$
-{m_an\over2\pi},
\efr
to the mass. As it stands, the second contribution in \twoconts\
is not well defined because the integral has a logarithmic divergence.
The problem is not due to any special nature of the soliton solution,
indeed it is to be expected, being due to the fact that we have not
renormalized the theory. We now pause to consider
this aspect in more detail.

Mercifully, the renormalization process in a two-dimensional field
theory is straightforward. The divergences can simply be removed by
normal-ordering the Hamiltonian. Working to lowest order in $\beta^2$
and introducing an ultra-violet cut-off $\Lambda$
$$
:\exp\,i\beta\alpha_j\cdot\phi:=\exp\,i\beta\alpha_j\cdot\phi\left[
1+\frac12\beta^2\sum_{a=1}^{n-1}{(\zeta_a\cdot\alpha_j)}^2\Delta_a+O(\beta^4)
\right],
\efr
where
$$
\Delta_a=\int_{-\Lambda}^\Lambda{dk\over4\pi}\,{1\over\sqrt{k^2
+m_a^2}},
\efr
and $\zeta_a$ is the $a^{\rm th}$ eigenvector of
$\sum_{j=1}^{n}\alpha_j\otimes\alpha_j$ of eigenvalue $(m_a/m)^2$.
Hence, to lowest order in $\beta^2$, the renormalized Hamiltonian is
equal to
$$
H_{\rm ren}={1\over2}\left({\partial\phi\over\partial t}\right)^2+
{1\over2}\left({\partial\phi\over\partial x}\right)^2
-{1\over\beta^2}\sum_{j=1}^{n}(m^2+\partial m^2_j)\left[\exp\left(
i\beta\alpha_j\cdot\phi\right)-1\right],
\efr
with
$$
\partial
m^2_j=\frac12(m\beta)^2\sum_{a=1}^{n-1}(\zeta_a\cdot\alpha_j)^2
\Delta_a.
\efr
The presence of the additional terms clearly changes the energy
of the soliton solution. These changes must be added to the quantum
correction to the soliton mass. It might be thought that we should now
consider the one-soliton solution for the renormalized equations of motion;
fortunately, this is unnecessary since the correction due to
change in the form of
the solution does not occur at the lowest order in $\beta^2$, precisely
because the soliton solution satisfies the ``bare'' equations of
motion \EQMOT.

Taking into account the counter-term from the renormalization, the
final expression for the mass correction for the $a^{\rm th}$ soliton
from the continuum solutions is
$$\eqalign{
\Delta M_a(2)=&-{m_an\over2\pi}+\sum_{b=1}^{n-1}\int_{-\Lambda}^\Lambda
{dk\over4\pi}\epsilon_b(k){d\rho_b(k)\over dk}\cr
 &\qquad\quad-
{1\over\beta^2}
\sum_{j=1}^n\int_{-\infty}^\infty dx\left[\exp(i\beta\alpha_j\cdot
\pb)-1\right]\partial m^2_j.\cr}
\nfr{intmass}

We now pause to calculate the integral
$$
\int_{-\infty}^\infty dx\,\left[\exp(i\beta\alpha_j\cdot\pb)-1\right].
\nfr{theint}
Using the explicit form for the stationary one-soliton solution, one finds
$$
\exp(i\beta\alpha_j\cdot\pb)-1=
-\left({m_a\over m}\right)^2\,{\omega^{aj}\exp(m_a
x+\xi)\over\left[1+\omega^{aj}\exp(m_ax+\xi)\right]^2}.
\efr
The integral \theint\ may now be calculated explicitly yielding
$$\left.
\left({m_a\over
m}\right)^2\,{1\over m_a\left[1+\omega^{aj}\exp(m_ax+\xi)\right]}\right\vert
_{-\infty}^\infty=-{m_a\over m^2}.
\efr

The important point about the resulting expression for the integral is
that it is independent of $j$. The contribution to the mass correction
\intmass\ from the renormalization counter-term can now be simplified:
$$
\eqalign{
-{1\over\beta^2}\sum_{j=1}^n\int_{-\infty}^\infty dx\,&\left[\exp(i\beta
\alpha_j\cdot\pb)-1\right]\partial m^2_j\cr
&=\frac12m_a\sum_{b=1}^{n-1}\sum_{j=1}^n(\zeta_b\cdot\alpha_j)^2\Delta_b\cr
&=\frac12m_a\sum_{b=1}^{n-1}\left({m_b\over m}\right)^2\int_{-\Lambda}^\Lambda
{dk\over4\pi}\,{1\over\sqrt{k^2+m_b^2}},\cr}
\efr
where we used the fact that $\zeta_b$ is an eigenvector of
$\sum_{j=1}^n\alpha_j\otimes\alpha_j$ of eigenvalue $(m_b/m)^2$.

The expression for the mass correction is now
$$\eqalign{
\Delta M_a(2)=-{m_an\over2\pi}&
+m_a\sum_{b=1}^{n-1}\int_{-\Lambda}
^\Lambda{dk\over4\pi}\left\{
\frac12\left({m_b\over
m}\right)^2\,{1\over\sqrt{k^2+m_b^2}}\right.\cr
&\left.-4\sqrt{k^2+m_b^2}\,{m_a^2+m_b^2-m_{a+b}^2
\over(m_a^2+m_b^2-m_{a+b}^2)^2+4m_a^2k^2}\right\}.\cr}
\efr
Using the fact that
$$
\sum_{b=1}^{n-1}(2m_a^2+2m_b^2-m_{a-b}^2-m_{a+b}^2)=2m_a^2n,
\efr
and
$$
\sum_{b=1}^{n-1}\left({m_b\over m}\right)^2={\rm
Tr}\left(\sum_{j=1}^n\alpha_j\otimes\alpha_j\right)=\sum_{j=1}^n\alpha_j
\cdot\alpha_j=2n,
\efr
one can easily verify that for large $|k|$ the integrand behaves as
$k^{-2}$ and so, as required, the logarithmic divergence is cancelled by the
renormalization counter-term. We can now let the
ultra-violet cut-off $\Lambda$ tend to infinity.

The remaining contribution is given in terms of a convergent integral:
$$
\Delta M_a(2)=m_a\left[-{n\over2\pi}+I(a,n)\right],
\efr
where
$$\eqalign{
I(a,n)=\int_{-\infty}^\infty {dk\over4\pi}\sum_{b=1}^{n-1}&
\left\{-4\sqrt{k^2+m_b^2}\,{m_a^2+m_b^2-m_{a+b}^2
\over(m_a^2+m_b^2-m_{a+b}^2)^2+4m_a^2k^2}\right.\cr
&\qquad\qquad\left.+\frac12\left({m_b\over
m}\right)^2\,{1\over\sqrt{k^2+m_b^2}}\right\}.\cr}
\efr
Unfortunately, we have not managed to evaluate the integral $I(a,n)$
analytically; except when $n=2$ in which case the integrand is
zero. However, by evaluating the integral numerically for a range of
$a$ and $n$ we find, to a very high degree of accuracy, that,
firstly, the integral is independent of $a$, and secondly it is
equal to the following functional form [\Ref{MRN}]:
$$
I(a,n)=\cases{-{1\over2}{\rm cosec}{\pi\over n}\qquad
&$n\in{\rm Odd}$\cr -{1\over2}{\rm cot}{\pi\over n}
&$n\in{\rm Even}$.\cr}
\efr

We can put the two contributions from the bound-states and the
continuum together to arrive at our main result:
$$
\Delta M_a=\Delta
M_a(1)+\Delta M_a(2)=m_a\left(-{n\over2\pi}+{1\over2}{\rm
cot}{\pi\over n}\right).
\nfr{MAIN}
Notice that the final result is valid for $n$ even {\it or\/} odd: the
asymmetry has cancelled out on adding the two contributions.

\chapter{Discussion}

We have succeeded in calculating the first quantum corrections to the
solitons masses in complex $sl(n)$ affine Toda field theories; and as
a by-product proving that the solitons are classically stable.
Although the calculations were somewhat lengthy the result is simple:
$$
\hat M_a=2nm_a\left[{1\over\beta^2}-{1\over4\pi}+{1\over4n}{\rm cot}
{\pi\over n}+O(\beta^2)\right],
\nfr{QSOLM}
where the first term is the classical mass. For the sine-Gordon theory
it is thought that the result is exact and there are no higher order
corrections; this example is recovered by setting $n=2$, giving
$$
\hat M={8m\over\beta^{\prime2}},
\efr
where $\beta'$ is the characteristic ``renormalized'' coupling
$$
\beta^{\prime2}={\beta^2\over1-\beta^2/4\pi}.
\nfr{SHIFT}
In the sine-Gordon theory it is well-known that the fundamental
particle is, in the quantum theory, a soliton anti-soliton
bound-state. However, it has been conjectured in ref. [\Ref{SM}]
that the same is true in the
more general theories. The soliton-soliton $S$-matrix written down in
ref. [\Ref{SM}] has poles corresponding to soliton bound states; in
particular there are $n-1$ distinct sets of scalar states,
carrying zero topological charge. The ground states of these sets, have
mass
$$
\hat m_a=2\hat M_a\sin{\pi\over n\lambda},\qquad a=1,2,\ldots,n-1,
\nfr{REL}
where $\hat M_a$ is the soliton mass and $\lambda$ is a coupling
constant which is known to be related to $\beta$ as [\Ref{SM}]
$$
\lambda={4\pi\over\beta^2}+\tilde\lambda,
\nfr{COUP}
where $\tilde\lambda$ is $O(1)$, but was not determined in ref. [\Ref{SM}].
By using \QSOLM\ and \COUP\ we can expand \REL\ to first order:
$$
\hat m_a=m_a\left[1+\left({1\over4n}{\rm cot}{\pi\over n}-{1\over4\pi}
-{\tilde\lambda\over4\pi}\right)\beta^2+O(\beta^4)\right].
\nfr{PART}
But \PART\ should be compared with the one-loop formula for the masses
of the fundamental particles in \QUANMASS\ (with
$\tilde\beta=i\beta$). They are consistent if $\tilde\lambda=-1$ so
$$
\lambda={4\pi\over\beta^{\prime2}},
\efr
to this order in $\beta$, where $\beta'$ is the shifted coupling
appearing in the sine-Gordon theory \SHIFT.

It would be interesting to investigate the question as to whether the
masses in \QSOLM\ are actually exact. On way to do that would be to
repeat the analysis of ref. [\Ref{wkb}], and calculate the corrections
to the masses of the breathers.

I would like to thank M.R. Niedermaier for finding the analytic
expressions for the integrals $I(a,n)$. I also have pleasure in thanking
J.M. Evans for contributing to the investigation of the classical
stability of the solitons in Section 2. Finally, I acknowledge the
support of Merton College, Oxford, through a Junior Research Fellowship.

\references

\beginref
\Rref{todasm}{A.E. Arinstein, V.A. Fateev and A.B. Zamolodchikov,
Phys. Lett. {\bf B87}  (1979) 389\newline
H.W. Braden, E. Corrigan, P.E. Dorey and R. Sasaki,
Nucl. Phys. {\bf B338} (1990) 689}
\Rref{wkb}{R.F. Dashen, B. Hasslacher and A. Neveu,
Phys. Rev. {\bf D11} (1975) 3424; Phys. Rev. {\bf D10} (1974) 4130}
\Rref{raj}{R. Rajaraman, Phys. Rep. {\bf 21C} (1975) 227}
\Rref{SOL}{T.J. Hollowood, `{\sl Solitons in affine Toda field
theories\/}', Oxford University preprint OUTP-92-04P, {\sl to appear
in\/}: Nucl. Phys. {\bf B}}
\Rref{GENSOL}{N.J. MacKay and W.A. McGhee, `{\sl Affine Toda solitons
and automorphisms of Dynkin diagrams\/}', Durham University and RIMS
preprint DTP-92-45, RIMS-890}
\Rref{SM}{T.J. Hollowood, `{\sl Quantizing SL(N) solitons and the
Hecke algebra\/}', Oxford University preprint OUTP-93-03P, {\sl to
appear in\/}: Int. J. Mod. Phys. {\bf A}}
\Rref{JON}{J.M. Evans, `{\sl Complex Toda theories and twisted reality
conditions\/}', Oxford University preprint OUTP-91-39P, {\sl to appear
in\/}: Nucl. Phys. {\bf B}}
\Rref{MRN}{M.R. Niedermaier, {\sl private communication\/}}
\endref
\ciao
